\documentclass[twocolumn,showpacs,preprintnumbers,amsmath,amssymb,superscriptaddress,10pt,aps,prb]{revtex4-2}
\usepackage{epsfig,amsopn}
\usepackage{graphicx}
\usepackage{epstopdf}
\usepackage{amsmath,amssymb}
\usepackage{amsthm}
\usepackage{enumerate}
\usepackage{cancel}
\usepackage{xcolor}
\usepackage[normalem]{ulem}
\usepackage[colorlinks=true,linktoc=page,linkcolor=magenta,citecolor=magenta]{hyperref}

\newcommand\bea{\begin{eqnarray}}
	\newcommand\eea{\end{eqnarray}}
\newcommand\beq{\begin{equation}}  
	\newcommand\eeq{\end{equation}}

\begin{document}

	\title{Two-channel Kondo problem in coupled interacting helical liquids}
	
	\author{Sourav Biswas}
	\affiliation{Department of Physics, Indian Institute of Technology - Kanpur, Kanpur 208 016, India.}
	
	\author{Alessandro De Martino}
	\affiliation{Department of Mathematics, City, University of London, EC1V 0HB London, UK}
	
	\author{Sumathi Rao}
	\affiliation{International Centre for Theoretical Sciences (ICTS-TIFR),
		Shivakote, Hesaraghatta Hobli, Bangalore 560089, India}
	
	\author{Arijit Kundu}
	\affiliation{Department of Physics, Indian Institute of Technology - Kanpur, Kanpur 208 016, India.}
	
	\begin{abstract}
		We study the two-channel Kondo problem in the context of two interacting helical liquids coupled to a spin-$\frac12$ magnetic impurity. We show that the interactions between the two helical liquids significantly affect the phase diagram and other observable properties. Using a multichannel Luttinger liquid formalism, we analyze both the Toulouse limit, where an exact solution is available, and the weak coupling limit, which can be studied via a perturbative renormalization group (RG) approach. We recover the  results for the `decoupled' limit (interactions between the helical liquids switched off) and point out deviations from the known results due to this coupling. The model under study is mapped to a model of two effectively decoupled helical liquids coupled to an impurity. The perturbative RG study shows that each of these channels can flow to either a  Ferromagnetic (FM) or an  Anti-Ferromagnetic (AFM) fixed point. We obtain  the phase diagram of the  coupled system  as a function of the system parameters.  The observable consequences of the interaction between the two channels are captured using linear response theory. We compute the negative correction to the conductance due to the Kondo scattering processes and show how it scales with the temperature as a function of inter-channel interaction. 
	\end{abstract}
	
	\date{\today} 
	
	\maketitle
	
	\section{Introduction}
	
	Topological systems have been at the center of research in condensed matter physics due to their exotic properties, one of them being the existence of topologically protected boundary modes~\cite{Haldane,Bernevig,Hasan}. In the case of two-dimensional topological systems, these modes are robust one-dimensional channels. The quantum spin Hall insulators, for example, host helical channels at the edge of the sample by virtue of time-reversal symmetry (TRS) of the bulk Hamiltonian~\cite{Kane1,Kane2,Bernevig2006}. For the purpose of our study, we specifically focus on one-dimensional helical channels present in two-dimensional topological systems. We emphasize that the low energy regime of these systems is spanned by the states representing helical channels of one-dimensional nature. 
	
	A peculiarity of one dimension is that, in spite of the presence of Coulomb interaction, the system remains exactly solvable, under certain conditions. It is well understood that the interacting physics of these edge modes is described by the Luttinger liquid (LL) theory~\cite{Zhang2006,Moore2006,Nagaosa2009}. The applicability of the LL formalism can be attributed to the linear dispersion of the edge states at low energies and to the topological protection against various back-scattering processes.  However, such systems may not be exactly solvable in the presence of impurities. Here, we are interested to study the effect of a single magnetic impurity on the one-dimensional helical channels formed at the boundary of two-dimensional topological systems, taking Coulomb interaction into account. The LL formed by the helical channel in the presence of Coulomb interaction is termed helical liquid (HL).
	
	It is well known that the Kondo effect describes the interaction between conduction electrons and a localized magnetic moment~\cite{Anderson1970,Nozieres1980,Nozieres1995,Coleman1995,Cox1998,bHewson}. This phenomenon can also be investigated when Coulomb interaction is present in the conduction channel~\cite{Furusaki1994,Furusaki2005}. So far, there have been studies addressing the problem of a spin-$\frac{1}{2}$ 
	magnetic impurity coupled to a single helical liquid~\cite{Zhang2009,Furusaki2011,Eriksson2012,Eriksson2013}, as well as to two helical liquids~\cite{Law2010,Posske2013,Lee2013,Posske2014}. 
	In particular, Posske et al.~\cite{Posske2013} have studied the problem of two helical liquids decoupled from each other and coupled to a magnetic impurity. They have considered the Toulouse limit, where the model is exactly solvable, and have analyzed the behavior of the Kondo screening cloud.
	
	
	Here, we study a model of two interacting helical liquids coupled to a spin-$\frac{1}{2}$ magnetic impurity, allowing for 
	forward scattering
	processes between the two helical liquids which preserve the symmetries 
	of the bare Hamiltonian. We show that the inclusion of all forward scattering processes allowed by symmetry modifies 
	the properties of the system and has observable consequences.
	The model can be mapped to a model of two decoupled channels interacting with the impurity.
	For a specific set of parameters values (the so-called Toulouse point)
	the mapping is essentially an Emery-Kivelson~\cite{Emery1992} mapping, which reduces 
	the interacting system to an exactly solvable two-channel resonant-level model~\cite{Emery1992,Toulouse1969,Zarand2000}.
	Away from the Toulouse point, the mapping still works, but the model is no longer solvable. 
	We then use perturbative renormalization group techniques to study the effects 
	of the Kondo interaction. 
	Perturbative RG techniques have been very instrumental in the study of Kondo effects~\cite{Zhang2006,Anderson1970,Nozieres1980}. One can use this technique to study the fixed points of the model, even though the model is not exactly solvable. One can further look into how these fixed points are modified as a function of system parameters. We note that previously the effect of scalar disorder has been studied in the context of a two channel Luttinger liquid set-up~\cite{Lerner2017_2,Lerner2017_1} using the multichannel Luttinger liquid (MLL) formalism. Our analysis extends the study of impurities  in a multichannel Luttinger liquid set-up to magnetic impurities as well. 
	
	The plan for the rest of the paper is as follows. 
	In Sec.~\ref{Mod}, we introduce the model of two interacting 
	helical liquids coupled to a spin-$\frac{1}{2}$ impurity. We diagonalize the 
	interaction terms (without the impurity) and, by using unitary transformations, 
	recast the coupling to the Kondo impurity into a simpler form.
	In Sec.~\ref{ExSol}, we focus on the Toulouse point, where the model can be reduced 
	to an exactly solvable one, and compute the impurity spectral function.
	In Sec.~\ref{weakcouplingRG}, we move away from 
	the exactly solvable point and use the perturbative renormalization group (RG) method to obtain the 
	Kondo temperatures for both channels and to study the fixed points as function of the system parameters. 
	In Sec.~\ref{ObsCon}, we present the explicit form of the correlation to the conductance as a function of the temperature. 
	Finally, in Sec.~\ref{Sum}, we present our conclusions and provide an outlook.
	Throughout this paper, we  set $\hbar=1$.
	\section{Model}\label{Mod}
	
	We consider a system of two interacting helical liquids coupled to a magnetic impurity expressed by the model Hamiltonian $H = H_{\rm LL} + H_{\rm K}$, where $H_{\rm LL}$ describes the bulk of the HLs and $H_{\rm K}$ represents a magnetic impurity coupled to 
	the HLs. 
	The bulk Hamiltonian takes the well-known form $H_{\rm LL}=H_0+H_{\rm int}$, 
	where $H_0$ is the bare part and $H_{\rm int}$ accounts for the Coulomb interaction present in the HLs. 
	We begin by writing the bare part of the model as~\cite{Kane1,Kane2,Zhang2006,Moore2006,Nagaosa2009}
	\begin{align}
		H_0
		&=-i\sum_{j,s} s v_j \int dx \Psi^{\dagger}_{j,s} \partial_x \Psi_{j,s},
	\end{align}
	where $\Psi_{js}$ are the field operators for the $j^{th}$ channel, with 
	$j \in \{1,2\}$, 
	and $v_j$ are the Fermi velocities. We assume that the right-moving modes, 
	denoted by $s=+$, carry spin up 
	and the left-moving modes, denoted by $s=-$, carry spin down. 
	
	\begin{figure}[t]
		\centering
		\includegraphics[width=.475\textwidth]{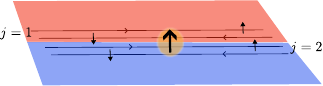}
		\caption{Schematic picture of the system under study. Two helical liquids, labeled by the index $j=1,2$, 
			propagate along the translational invariant direction $\hat{x}$. The right-moving fermions carry a spin up ($\uparrow$) index and the left-moving fermions a spin down ($\downarrow$) index. The two helical liquids are in close proximity to each other and are coupled to a spin-$\frac{1}{2}$ impurity, located at $x=0$.} \label{fig:setup}
	\end{figure}
	Next, we write down the interacting part of the model arising from the (screened) Coulomb interaction. We allow forward scattering processes between the two channels, which we write as
	\begin{align}
		\label{Hint}
		H_{\rm int}
		&= \int dx \Big[ g^{(1)}_4 ( \rho^2_{1R} + \rho^2_{1L} ) + 2g^{(1)}_2 \rho_{1R}\rho_{1L}  + g^{(2)}_4 ( \rho^2_{2R} \nonumber \\  & + \rho^2_{2L} )  + 2g^{(2)}_2 \rho_{2R}\rho_{2L}  + 2g^{(12)}_4 ( \rho_{1R}\rho_{2R} + \rho_{1L}\rho_{2L} ) \nonumber \\ 
		& + 2g^{(12)}_2 ( \rho_{1R}\rho_{2L} + \rho_{1L}\rho_{2R} ) \Big] ,
	\end{align}
	where $\rho_{j,s} = \Psi^{\dagger}_{j,s}
	\Psi_{j,s} $ is the fermionic density operator. Here $g^{\zeta}_{2},g^{\zeta}_{4}$ follow 
	the standard $g$-ology convention with $g^{\zeta}_2$  denoting forward scattering processes involving density operators of movers in opposite directions and $g^{\zeta}_4$ denoting processes with movers in the same direction. The superscript $\zeta=1,2$ denotes scattering within individual channels $j=1,2$, whereas  $\zeta=12$ denotes scattering involving both channels. We note that the bare Hamiltonian $H_0$ is topologically protected and the terms included in Eq.~\eqref{Hint} are allowed by TRS. In addition, these interacting liquids are coupled to a spin-$\frac12$ impurity, with the Kondo Hamiltonian given by
	\begin{align}
		H_{\rm K} 
		&= \sum_{j,s}  J_{z,j} s\Psi^{\dagger}_{j,s}(0) \Psi_{j,s}(0) \sigma^{z} \nonumber \\
		&+ \sum_{j=1,2}  J_{\perp,j} \Big( \Psi^{\dagger}_{j,+}(0) \Psi_{j,-}(0) \sigma^{-} + {\rm h.c.} \Big),
	\end{align}	
	where $J_{z,j}$ and $J_{\perp,j}$ are the Kondo couplings for the $j^{th}$ channel, 
	and $\sigma^z$, $\sigma^{\pm} = \sigma^x \pm i \sigma^y$ are the spin-$\frac{1}{2}$ 
	operators for the impurity located at $x=0$. 
	
	To proceed further, we employ the bosonization technique~\cite{bGiamarchi,bFradkin,bGogolin}.
	Since an HL has the same number of degrees of freedom as a spinless LL~\cite{Zhang2009},
	two 
	bosonic fields $\phi_j,\theta_j$ are sufficient to bosonize the Hamiltonian $H$. 
	We bosonize the fermion operator 
	using the identity~\cite{bGiamarchi,bFradkin,bGogolin}
	\begin{equation}
		\Psi_{j,s} = (2 \pi \xi_j)^{-1/2} e^{-i\sqrt{\pi}(\theta_j - s \phi_j)}, 
	\end{equation}
	where $\xi_j$  is a microscopic cut-off length for channel $j$. 
		The Klein factors have been neglected since they always appear in pairs in the quantities of interest we compute.
	We note that $\rho_{j,s}= \frac{1}{2\sqrt{\pi}}\partial_x (\phi_j-s \theta_j)$ and $\Pi_j = \partial_x \theta_j$. 
	By combining the bosonized $H_0$ and $H_{\rm int}$, we arrive at the multichannel Luttinger liquid (MLL) Hamiltonian 
	\begin{align}
		\label{Hd_para}
		H_{\rm LL} & = H_0 + H_{\rm int} \nonumber \\
		&= \frac{1}{2} \int dx \Big(  \partial_x \Phi^{T}M_{\phi} \partial_x \Phi 
		+ \partial_x \Theta^{T}M_{\theta}  \partial_x \Theta  \Big) ,\
	\end{align}	
	where we  have used the notation $\Phi =( \phi_1 ~~\phi_2 )^T$, $\Theta =( \theta_1 ~~\theta_2 )^{T}$ and
	\begin{align}\label{Mpt}
		M^{ij}_{\phi} &=  (v_i  + \frac{g^{(i)}_4+g^{(i)}_2}{\pi}) \delta_{ij} +\frac{g^{(12)}_4+g^{(12)}_2}{\pi}(1-\delta_{ij}), \\
		M^{ij}_{\theta} &=  (v_i  + \frac{g^{(i)}_4-g^{(i)}_2}{\pi}) \delta_{ij} +\frac{g^{(12)}_4-g^{(12)}_2}{\pi}(1-\delta_{ij}). 
	\end{align}
	We can now diagonalize $H_{\rm LL}$ using standard methods (see, e.g.,~\cite{Yurkevich2013,Yurkevich2021}). 
	We assume that  $H_{\rm LL}$ is diagonal in $\tilde \Theta$ and $\tilde \Phi$ fields, where $\tilde{\Phi }=( \tilde{\phi}_1 ~~\tilde{\phi}_2 )^T$, $\tilde{\Theta} =( \tilde{\theta}_1 ~~\tilde{\theta}_2 )^{T}$ and $\tilde{\Pi}_j = \partial_x \tilde{\theta}_j$.  These fields are related to the  $\Phi$ and $\Theta$ fields via  linear transformations  $\Phi =V_\phi \tilde \Phi$ and $\Theta =V_\theta \tilde\Theta$ such that $\tilde \Theta^T \tilde \Phi=\Theta^T  \Phi$.  The explicit forms of $V_\phi$ and $V_\theta$ are  found to be
	\begin{align}
		V_\phi &= U^T_\phi D^{-\frac{1}{2}}_\phi \mathcal U^T \mathcal D^\frac14, \\
		V_\theta &= U^T_\phi D^\frac12_\phi \mathcal U^T \mathcal D^{-\frac14},
	\end{align}
	where $U_\phi$ is a matrix that diagonalizes $M_\phi$ of Eq.~\eqref{Mpt}, $D_\phi$ is a diagonal matrix with the eigenvalues of $M_\phi$ as its diagonal entries. The  orthogonal matrix $\mathcal U$ and the diagonal matrix $\mathcal D$ are obtained from the product of matrices $ D^{\frac12}_\phi U_\phi M_\theta U^T_\phi D^\frac12_\phi$ by diagonalizing as $D^{\frac12}_\phi U_\phi M_\theta U^T_\phi D^\frac12_\phi = \mathcal{U}^T \mathcal{D}\mathcal{U}$. This procedure enables us to write the two-channel Luttinger liquid Hamiltonian $H_{\rm LL}=H_0+H_{\rm int}$ as
	\begin{equation}\label{HLdc}
		H_{\rm LL}  = 
		\sum_{j=1,2} \frac{u_{j}}{2} \int dx 
		\left[ (\partial_x \tilde{\theta}_{j})^{2}  + (\partial_x \tilde{\phi}_{j})^{2}  \right] , 
	\end{equation}
	where the renormalized velocities $u_j$ are the diagonal entries of $\mathcal{D}^{\frac12}$. The Kondo Hamiltonian, in terms of the new fields, takes the form
	\begin{align}
		H_{\rm K} 
		&=\sum_{j=1,2} \left[ 
		-\frac{ \tilde J_{z,j} }{\sqrt{\pi}}  \tilde{\Pi}_j(0) \sigma^{z} \right. \nonumber \\
		&\left. + \frac{ J_{\perp,j} }{2 \pi \xi_j}  \left( e^{i 2 \sqrt{\pi} [V^{j1}_{\phi} \tilde{\phi}_1(0) + V^{j2}_{\phi}\tilde{\phi}_2(0) ]} \sigma^{+} 
		+ {\rm h.c.} \right)  \right] ,\label{HKcoupled}
	\end{align}	
	where
	\begin{equation}
		\label{rotJ}
		\tilde J_{z,j} = \sum_{k=1,2}J_{z,k} V_{\theta}^{kj}.
	\end{equation} 
	
	At this point, we make a short digression to understand the decoupled limit from the calculations done so far. We notice that if we neglect the inter-channel interactions, $g^{(12)}_{2,4}=0$, then $M_{\phi,\theta}$ are diagonal. Hence $V_\phi$ and $V_\theta$ are also diagonal and given by $V^{ij}_\phi=\sqrt{K_j}\delta_{ij}$ and $V^{ij}_\theta=\delta_{ij}/\sqrt{K_j}$,
	where $$K_j =\sqrt{1+\frac{g^{(j)}_4-g^{(j)}_2}{\pi v_j}} \Big/ \sqrt{1+\frac{g^{(j)}_4+g^{(j)}_2}{\pi v_j}}$$ is the usual Luttinger
	liquid parameter for channel $j$. Therefore, $ \tilde \phi_j  =  \phi_j/\sqrt{K_j}$,
	$\tilde \Pi_j= \sqrt{K_j} \Pi_j$, $u_j=v_j/K_j$, and $\tilde J_{z,j}=J_{z,j}/\sqrt{K_j}$, and we recover the Hamiltonian considered in \cite{Law2010,Posske2013}. 
	
	The Hamiltonian~\eqref{HLdc} describes two effectively decoupled HLs obtained by the diagonalization procedure when $g^{(12)}_{2,4} \neq 0$. The new decoupled fields $\tilde{\phi}_j$ have been used to rewrite the Kondo Hamiltonian. 
	We observe from Eq.~\eqref{HKcoupled} that both fields $\tilde{\phi}_1$ and $\tilde{\phi}_2$ appear in each of the exponential terms of $H_{\rm K}$. This is a manifestation of the finite inter-channel scattering processes $g^{(12)}_{2,4}$. Hence, even if $H_{\rm LL}$ can be cast into a diagonal form, 
	the Kondo Hamiltonian still couples the two fields.
	We proceed further to reduce the full Hamiltonian $H$ to a Hamiltonian describing two decoupled interacting channels coupled to a single Kondo impurity, by devising a unitary transformation $U_{\rm d} = e^{i 2\sqrt{\pi} (\lambda_1 \tilde{\phi}_1(0)	+ \lambda_2 \tilde{\phi}_2(0) )\sigma^z}$ and choosing $\lambda_{1,2}$ appropriately to arrive at $\tilde{H} \equiv U_{\rm d} H U^{\dagger}_{\rm d}$ given by
	\begin{align}
		\label{eq:H_2CK_ind}
		\tilde{H} 
		&=\sum_{j=1,2} \left[ 
		\frac{u_j}{2}\int dx \left[  \tilde \Pi_j^2 +  ( \partial_x \tilde \phi_j )^2 \right] \right. \nonumber \\
		&\left. -\frac{\tilde J'_{z,j}}{\sqrt{\pi}} \tilde \Pi_j(0) \sigma^{z} 
		+ \frac{J_{\perp,j} }{2 \pi \xi_j} \left(
		e^{i 2 \sqrt{\pi} \kappa_j \tilde \phi_j(0)} \sigma^{+} + {\rm h.c.}  \right)
		\right],
	\end{align}	
	where 
	\begin{equation}
		\label{defkappa}
		\kappa_j = V^{jj}_{\phi}-V^{\bar jj}_{\phi} \quad {\rm and}\quad   \tilde J'_{z,j} = \tilde J_{z,j} - 2\pi u_j V_\phi^{\bar j j}.
	\end{equation}
	Here, $\bar j=2,1$ for $j=1,2$. We refer to  App.~\ref{app-DCh} for the details of the derivation. 
	We use this Hamiltonian in Sec.~\ref{weakcouplingRG} to derive the RG flow of the Kondo couplings. 
	
	Alternatively, we can use the unitary transformation to cancel the $\tilde J_{z,j}$-terms. 
	This is accomplished by setting $\lambda_j = -\frac{\tilde J_{z,j}}{2\pi u_j}$, as shown in App.~\ref{app-DCh}. 
	We then arrive at 
	\begin{align}
		\label{Ham}
		\tilde{H} 
		& = \sum_{j=1,2}\left[ 
		\frac{u_j}{2} \int dx \left[  \tilde{\Pi}_j^2 +  ( \partial_x \tilde{\phi}_j )^2  \right] \right. \nonumber \\
		&\left. + \frac{J_{\perp,j} }{2 \pi \xi_j} \left(
		e^{i 2 \sqrt{\pi} \sum_k \kappa_{jk}\tilde{\phi}_k(0)} \sigma^{+} + {\rm h.c. }\right) \right],
	\end{align}	
	where we have defined 
	\begin{equation}
		\kappa_{jk}=V_\phi^{jk} - \frac{\tilde J_{z,k}}{2\pi u_k}. \label{kap}
	\end{equation}
	This Hamiltonian is the starting point for the calculation of observables. In the next section, we study a particular limit in which $H = H_{\rm LL} + H_{\rm K}$ is exactly solvable, and in Sec.~\ref{weakcouplingRG} we use the perturbative RG approach to study the weak coupling limit beyond the solvable point.
	
	Before moving on, we briefly comment on the nonlinear terms that have been omitted in the interacting Hamiltonian. 
	In general, the inclusion of interaction-induced backscattering operators 
	can open a gap and render the gapless Luttinger liquid physics invalid. 
	However, a helical liquid is topologically protected against intra-channel backscattering terms.
	We consider here a regime in which also the inter-channel 
	backscattering terms are negligible~\cite{Nagaosa2009}.
	Furthermore, we assume that the system is away from half-filling
	and neglect all Umklapp processes~\cite{Nagaosa2009}, as further discussed in App.~\ref{app-BSDim}.
	In the same appendix, we also show that the  Kondo spin-flip scattering involving two different channels becomes irrelevant under certain conditions imposed on the system parameters. We only take into account the Kondo spin-flip scatterings within the same channel, assuming that the aforementioned irrelevance is satisfied.
	

	\section{Exactly Solvable Point} 
	\label{ExSol}
	
	In this section, we show that for special values of the system parameters the model admits an exact solution~\cite{Emery1992,Sengupta1994}. In fact, for the Hamiltonian~\eqref{Ham}, 
	there are two possible sets of conditions under which the mapping can be achieved. 
	The first set of conditions is
	\begin{align}
		\label{SP1}
		\kappa_{11}=\kappa_{22}=0, \quad \quad \kappa^2_{12}=\kappa^2_{21}=\frac12.  
	\end{align}
	The first condition ensures that the two channels decouple, i.e., 
	only one field appears in each exponential operator in the Kondo interaction. 
	The second ensures that the Kondo interaction has the scaling 
	dimension of a fermionic field. Using Eq.~\eqref{kap}, this set of conditions can be written in terms of  
	$V_\phi$ and $\tilde J_{z,j}$ as
	\begin{align}
		& \tilde J_{z,1} = 2\pi u_1 V_\phi^{11}, \quad  \tilde J_{z,2} = 2\pi u_2 V_\phi^{22},   \\ 
		&(V_\phi^{11}-V_\phi^{21})^2 = (V_\phi^{22}-V_\phi^{12})^2 = \frac12. \label{condition}
	\end{align}
	These conditions can only be satisfied when $V_\phi$ is a non-diagonal matrix. When these conditions hold, 
	the two-channel Kondo problem can be 
	mapped to a resonant-level problem~\cite{Toulouse1969,Emery1992,Zarand2000}. 
	We can find out the solution in terms of the system parameters, using the definition of $V_\phi$.
	
\begin{figure}[h!]
		\centering
		\includegraphics[width=.325\textwidth]{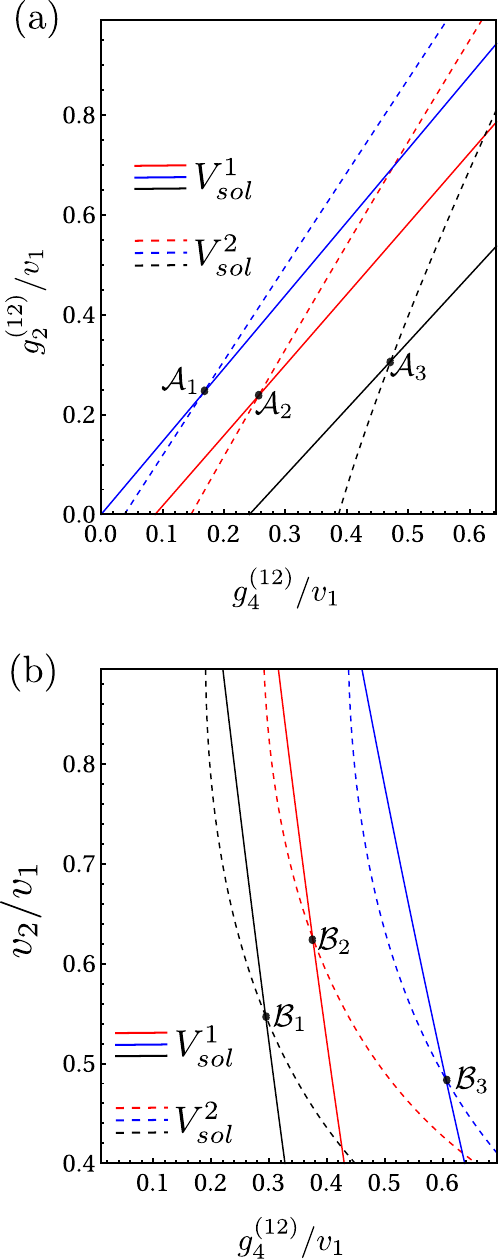}
		\caption{Two lines given by $V^{1}_{sol}=0$ and $V^{2}_{sol}=0$ are plotted as a function of system parameters. The intersection of these two lines denotes an exactly solvable point, where the model can be transformed into a model of two non-interacting channels coupled to a single impurity. These intersections are denoted by separate symbols $\mathcal{A}_j$ and $\mathcal{B}_j$. Changing the system parameters shifts the exact solution. In (a) we have varied $g^{(12)}_{2}$ and $g^{(12)}_{4}$. The origin corresponds to the decoupled limit of the problem. In (b),  $v_2$ and $g^{(12)}_{4}$ are varied. All the parameters are scaled by $v_1$. The other parameters which are required to generate this plot are listed in Table.~\ref{Tab1}.  }\label{fig:Exsol}
	\end{figure}

	Note that the alternative choice
	\begin{align}
		\label{SP2}
		\kappa_{12}=\kappa_{21}=0, \quad \quad \kappa^2_{11}=\kappa^2_{22}=\frac12
	\end{align}
	becomes
	\begin{align}
		& \tilde J_{z,1} = 2\pi u_1 V_\phi^{21}, \quad  \tilde J_{z,2} = 2\pi u_2 V_\phi^{12}, \\ 
		&(V_\phi^{11}-V_\phi^{21})^2 = (V_\phi^{22}-V_\phi^{12})^2 = \frac12. \label{condition2}
	\end{align}
	We see that Eq.~\eqref{condition2} is precisely the same as Eq.~\eqref{condition}. Both choices lead to the reduction of Eq.~\eqref{Ham} to a resonant-level model with two non-interacting channels coupled to a single magnetic impurity. Since the exact solution depends on relationships between different parameters of the model, it is clear that as the parameter values change, it is only for certain values that we will get an exact solution. In Fig.~\ref{fig:Exsol} we plot the two conditions $(V_\phi^{11}-V_\phi^{21})^2 -\frac{1}{2} \equiv V_{sol}^1=0$ and $(V_\phi^{22}-V_\phi^{12})^2-\frac12 \equiv V_{sol}^2=0$ as a function of two parameters at a time  (the other parameters have been fixed at the values given in Table I).  Then, the 
	intersections of the curves $V_{sol}^1=0$ and $V_{sol}^2=0$ give the points where an exact solution is possible.
	As discussed in the previous section, one has to choose the parameters in a way such that Eq.~\eqref{Ham} is sufficient to describe all possible scattering processes. Following the analysis done in App.~\ref{app-BSDim}, we make the choice of the parameters in a way such that
	\begin{align}\label{Cond_drop}
		\frac{1}{4} \sum_{j=1,2} \left(V_\theta^{1j}-V_\theta^{2j} \right)^2 
		+ \frac{1}{4} \sum_{j=1,2} \left(V_\phi^{1j}
		\mp V_\phi^{2j}\right)^2 > 1.
	\end{align}
	
	\begin{table}[htb] 
		\centering
		\begin{tabular}{|c|c|c|c|c|c|c|c|}
			\hline
			&$v_2/v_1$ & $g^{(1)}_2/v_1$ & $g^{(2)}_2/v_1$ & $g^{(1)}_4/v_1$ & $g^{(2)}_4/v_1$ & $g^{(12)}_2/v_1$ & $g^{(12)}_4/v_1$ \\
			\hline 
			$\mathcal{A}_1$ & 1.5 & 2.0 $\pi$ & 2.4 $\pi$ & 2.2$\pi$ & 2.5$\pi$ & $\dots$ & $\dots$ \\
			\hline
			$\mathcal{A}_2$ & 0.5 & 1.25$\pi$ & 1.4$\pi$ & 1.4$\pi$ & 1.5$\pi$ & $\dots$ & $\dots$ \\
			\hline
			$\mathcal{A}_3$ & 0.4 & 1.2$\pi$ & 1.4$\pi$ & 1.6$\pi$ & 1.5$\pi$ & $\dots$ & $\dots$ \\
			\hline 
			$\mathcal{B}_1$ & $\dots$ & 1.45$\pi$ & 1.475$\pi$ & 1.6$\pi$ & 1.6$\pi$ & 0.35 & $\dots$ \\
			\hline
			$\mathcal{B}_2$ & $\dots$ & 1.45$\pi$ & 1.475$\pi$ & 1.5$\pi$ & 1.5$\pi$ & 0.5 & $\dots$ \\
			\hline
			$\mathcal{B}_3$ & $\dots$ & 1.45$\pi$ & 1.65$\pi$ & 1.6$\pi$ & 1.75$\pi$ & 0.75 & $\dots$ \\
			\hline
		\end{tabular}
		\caption{Values of the system parameters used in Fig.~\ref{fig:Exsol}. The blank boxes with dots
			stand for the parameters which are varied in the plots.} \label{Tab1}
	\end{table}
	The choice of the parameters for the mapping to the resonant level model has been made in a way such that the vertex operator, $e^{i 2 \sqrt{\pi} \sum_k \kappa_{jk}\tilde{\phi}_k(0)}$  can be refermionized using the same bosonization identity that  we  used earlier so that the model reduces to that of free fermions. We can write the non-chiral boson fields used for describing the HL as $\tilde{\phi}_j = (\tilde{\phi}_{j,L} + \tilde{\phi}_{j,R} )$ and $\tilde{\theta}_j = (\tilde{\phi}_{j,L} - \tilde{\phi}_{j,R} )$ where ${R}$ and ${L}$ denote the right and left movers. In order to cast the Hamiltonian as an exactly solvable non-interacting fermion model, we first focus on the Luttinger liquid defined on the positive and negative $x$-axis, separately. The system on the right and left half-lines is ``unfolded"~\cite{Affleck1995,Emery1992,bGiamarchi} so that the Hamiltonian is presented in a chiral form. We follow the convention of writing $\tilde{\phi}_{j,R}(x)$ and $\tilde{\phi}_{j,L}(x)$ fields defined on the positive $x$-axis in terms of $\tilde{\phi}^e_{j,R}(x)$ and $\tilde{\phi}^e_{j,R}(-x)$ fields defined on the full $x$-axis. This is done as follows: $\tilde{\phi}^e_{j,R}(x) = \tilde{\phi}_{j,R}(x)$ and $\tilde{\phi}^e_{j,R}(-x) = \tilde{\phi}_{j,L}(x)$. One can obtain two chiral liquids in the bulk of the channel using these identities. Then, one must identify the chiral boson fields to be equal at $x=0$ i.e. $\tilde{\phi}^e_{j,L}(0) = \tilde{\phi}^e_{j,R}(0)$. We notice that the Kondo scattering term is defined only at $x=0$. We choose one chiral field out of $\tilde{\phi}^e_{j,R}(0)$ and $\tilde{\phi}^e_{j,L}(0)$, to write the vertex operator of the Kondo interaction term. After doing so, the Hamiltonian can again be written on the full line in terms of chiral fields only. However, depending on the choice of the chiral field used to write down the vertex operator in the Kondo scattering term, one chiral field remains decoupled from the impurity and can be discarded from the resonant level model. In what follows, we use the right-moving bosonic field to write down the exact solution. The bosonization identity $ \Psi_{j} = (2 \pi \xi_j)^{-1/2} e^{i2\sqrt{\pi}\tilde{\phi}^e_{j,R}}$  can be used to write the model in terms of fermions. In this limit, Eq.~\eqref{Ham} can be expressed in terms of chiral spinless fermions $\Psi_j(x)$ \cite{Furusaki2011} as
	\begin{align} 
		\label{HT}
		H_{\rm T} & = \sum_j \left[ -i u_j \int dx \Psi_j^{\dagger}(x) \partial_x \Psi_j(x) 
		+ \epsilon_d d^{\dagger}d  \right. \nonumber \\
		& \left.
		+ \frac{J_{\perp,j} }{\sqrt{2 \pi \xi_j}} 
		\left( d^{\dagger} \Psi_j(0) + \Psi^{\dagger}_j(0) d \right) \right] ,
	\end{align}	
	where the impurity spin residing at $x=0$ has been modeled by a discrete level, such that $\sigma_z = d^{\dagger}d-1/2 $. The operator $ d^{\dagger} $ creates a spinless fermion in the discrete level and $\epsilon_d $ is the chemical potential at the site of the discrete level. We refer to App.~\ref{app-2RLM} for the details of the exact solution. At the exactly solvable point, one can self-consistently compute the energy spectrum and the impurity spectral function~\cite{bColeman}. In our case, the spectral function turns out to be a Lorentzian with a level width $\Gamma = \sum_j \frac{J^2_{\perp,j} }{4 \pi \xi_j u_j}$. We see that the two contributions coming from the two independent channels add up directly.
	
	In the decoupled limit (which can be obtained by switching off the interchannel forward scattering 
	processes $g^{(12)}_{2,4}$), our result matches with one of the exactly solvable points derived in~\cite{Posske2013}. In this limit, $M_{\phi,\theta}$ are diagonal and $K_j = \kappa^2_j =(V^{jj}_{\phi})^2$. We have chosen $\kappa_j = 1/\sqrt{2}$ and as a result $K_1+K_2=1$. We note that our exact solution does not require the channels to have equal velocities. In fact, due to the presence of off-diagonal terms in $M_{\phi,\theta}$, the renormalized velocities of the diagonalized Hamiltonian are not equal in our case. At the exactly solvable point choosing $K_j = 1$ leads to an effectively non-interacting fermionic model coupled to magnetic impurity, if inter-channel processes are switched off. This particular limit is not of our interest. However we note, this would lead to another exactly solvable limit derived in \cite{Posske2013}.
	
	\section{Beyond The Exactly solvable limit}
	\label{weakcouplingRG}
	
	\begin{figure}[t!]
		\centering
		\includegraphics[width=.375\textwidth]{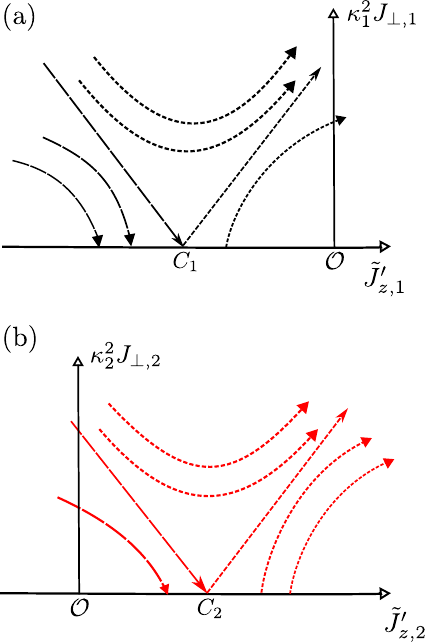}
		\caption{Schematic representation of the RG flow in Eqs.~\eqref{RG_KT_1} and~\eqref{RG_KT_2}. In panel (a)
			and (b) we show two sets of flow trajectories pertaining to the two different channels.  
			The symbol $\mathcal{O}$ denotes the point $(J_{\perp,j},\tilde J'_{z,j})=(0,0)$.
			The position of the starting point of the trajectory relative to $C_j$
			(rather than to $\mathcal{O}$) determines whether the system flows to a FM or an AFM fixed point.} 
		\label{fig:RG}
	\end{figure}
	
	In this section, we use the perturbative renormalization group technique 
	to analyze the flow of the Kondo couplings~\cite{maciejko2012,bAltland}. 
	By using the effectively decoupled Hamiltonian in Eq.~\eqref{eq:H_2CK_ind}, 
	we find that for each channel $j$ the RG equations (up to second order
	in the couplings) are given by
	\begin{align}
		\label{eq:RG_1}
		\frac{d \tilde J'_{z,j}}{dl} & = 
		\nu_j \kappa^3_j J^2_{\perp,j}, \\ \label{eq:RG_2}
		\frac{d J_{\perp,j}}{dl} &=
		(1-\kappa^2_j) J_{\perp,j} +
		\nu_j  \kappa_j \tilde J'_{z,j} J_{\perp,j} ,
	\end{align}	
	where $\nu_j \equiv \frac{ 1 }{ \pi u_j}$ 
	\footnote{Note that the RG equations appear to differ from those in Ref.~\cite{maciejko2012}. 
		However, the apparent mismatch is explained when we take into account the difference 
		in the  definition of $J_{z,j}$. $J_{z,j}$ here is equivalent to $J_{z,M}K_j$, 
		where $J_{z,M}$ is the coupling defined in Ref.~\cite{maciejko2012}.}. 
	We emphasize that the MLL formalism is instrumental in mapping the results to a form 
	similar to the known one-channel counterpart~\cite{Zhang2006,Zhang2009,maciejko2012}. 
	The details of the derivation are provided in the App.~\ref{app-RG}. 
	The equations~\eqref{eq:RG_1} and~\eqref{eq:RG_2} can be put in a more compact form 
	by defining $\nu_j \bar J_{z,j} = \nu_j \kappa_j \tilde J'_{z,j} + 1 - \kappa^2_j$ as
	\begin{align} \label{RG_KT_1}
		\frac{d \bar J_{z,j}}{dl} &= \nu_j \kappa_j^4 J^2_{\perp,j}, \\ \label{RG_KT_2}
		\frac{d J_{\perp,j}}{dl}  &= \nu_j  \bar J_{z,j} J_{\perp,j}.
	\end{align}
	The trajectories of the flow equations are given by $(\kappa^2_j J_{\perp,j})^2 - ( \bar{J}_{z,j})^2=c$, where $c$ is a constant. It is known that a single channel either flows to an anti-ferromagnetic (AFM) or a ferromagnetic (FM) fixed point (FP)~\cite{Anderson1970,Zhang2009}. Earlier works have shown how the RG flow in such  a system depends on the Luttinger parameter~\cite{Zhang2009,Zhang2006}. In our case, each of the effectively decoupled channels behaves like a single Luttinger liquid wire and can flow to either FM- or AFM-FP separately. However, the inclusion of inter-channel interactions in Eq.~\eqref{Hint} modifies the range of parameters of the system for which the couplings flow either to the FM or the AFM fixed point. 
	
	We plot a schematic flow diagram pertaining to the above RG equations in Fig.~\ref{fig:RG}. If the Kondo couplings of channel $j$ of Eq.~\eqref{eq:H_2CK_ind} flow to infinity, then the impurity is strongly coupled to channel $j$. On the other hand, if $J_{\perp,j}$ renormalizes to zero, then the impurity is weakly coupled with channel $j$. The transition point $C_j$ separates the $\tilde{J}'_{z,j}$ axis into two portions having opposite flows. To study different FPs, we choose $J_{\perp,j}=0$ of Fig.~\ref{fig:RG} as the initial condition. On this line, whether the system flows to FM or AFM fixed point is decided by the position of $C_{j}$.
	\begin{figure}[]
		\centering 
		\includegraphics[width=0.325\textwidth]{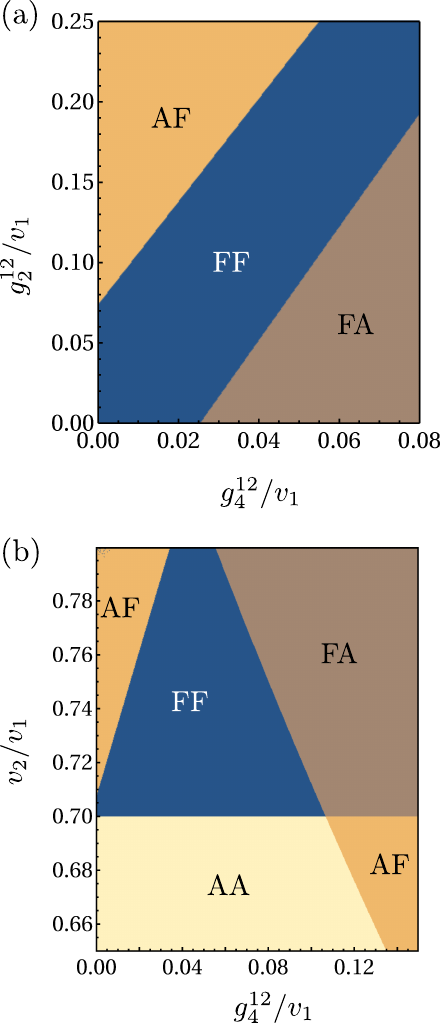}
		\caption{The above diagrams show the different FPs to which the  two channels flow,  
			as a function of the system parameters. We refer to these diagrams of the parameter space as phase diagrams. Each of the effectively decoupled channels 
			can flow to either the $\rm A$ or the $\rm F$ fixed point, starting from $J_{\perp,j}=0$, 
			(see the discussion in the main text). 
			In panel (a) we set $v_2/v_1=0.75$, $g^{(1)}_2/v_1= 0.65 \pi$, $g^{(2)}_2/v_1= 0.7 \pi$, 
			$g^{(1)}_4/v_1= 1.275 \pi$, $g^{(2)}_4/v_1= 1.5\pi$. In panel (b) we set 
			$g^{(1)}_2/v_1= 0.5\pi$, $g^{(2)}_2/v_1= 0.6\pi$, $g^{(1)}_4/v_1= 1.2\pi$, 
			$g^{(2)}_4/v_1= 1.4\pi$, $g^{(12)}_2/v_1= 0.175$.}
	\label{fig:Phase}
\end{figure} 

In Fig.~\ref{fig:RG}, the position of the transition point $C_j$ on $\tilde{J}'_{z,j}$ axis depends on the solution of $\nu_j \kappa_j \tilde J'_{z,j} + 1 - \kappa^2_j=0$. We know from Eq.~\eqref{defkappa} that $\kappa_j = V^{jj}_{\phi}-V^{\bar jj}_{\phi}$. Hence, we see that inter-channel interactions  shift the position of the transition point on $\tilde J'_{z,j} $ axis. As long as $\kappa_j \neq 1$ the transition point does not lie at $\tilde J'_{z,j} = 0$ which we denote by $\mathcal{O}$. 

Next, we look into the RG flow of each of the effectively decoupled channels of Eq.~\eqref{eq:H_2CK_ind}. 
It is easy to identify, from Fig.~\ref{fig:RG}(b), that for $\kappa_2 > \kappa_2^{-1}$ there is a region on the positive $\tilde J'_{z,2}$ axis between $C_2$ and $\mathcal{O}$, where even if the Kondo coupling is positive  i.e. AFM like, the system flows to an FM fixed point. We denote these fixed points by F. By the same token, a channel can flow to an  AFM FP despite being expected to flow to an FM FP. Such FPs are denoted by A. For example in Fig.~\ref{fig:RG}(a), the transition point  $C_1$ lies to the left of $\mathcal{O}$ and FPs of type A are obtained for the choice $\kappa_1 < \kappa_1^{-1}$. We can combine these FPs of both channels and name them `$ \mathcal{P}_j \mathcal{P}_{j'}$', where $ \mathcal{P}_j$ denotes the FP of channel $j$ and can be either $\rm A$ or, $\rm F$. Once we have identified the fixed points we can study their dependence on different parameters, as shown in Fig.~\ref{fig:Phase}. The phase diagrams in Fig.~\ref{fig:Phase} are representations of the parameter space of the Hamiltonian, where different fixed points are reached in different segments of the parameter space. These parameters have to be chosen in a way such that conditions written down in Eq.~\eqref{Cond_drop} are satisfied. We note that a similar analysis can also be done for different choices of $J_{\perp,j}$, as an initial condition. 

The characteristic feature of an FM FP is the renormalization of $J_{\perp, j}$ to zero. We note that the spin-flip scattering processes are governed by this coupling and as a result, in an FM FP no strongly coupled bound state formation takes place between the conduction electron and the impurity spin. Another way to understand this is that, if  $J_{\perp, j}$ flows to zero then the conduction electrons of channel $j$ do not take part in spin-flip Kondo scattering processes. However, if $J_{\perp, j}$ flows to infinity, then the electrons of the effectively decoupled channel ($j$) participate in spin-flip scattering processes leading to the formation of a bound state between the electron of channel $j$ and the impurity spin. This is a feature of AFM FP. Thus, in either AF or FA phase, there is only one channel that contributes to bound state formation. This bound state is a spin singlet and in these two phases, the impurity is screened. In the FF phase, none of the channels are strongly coupled with the impurity and hence there is no screening. In the AA phase, both channels try to couple anti-ferromagnetically and as a consequence over-screening of the impurity can take place. One should keep in mind that the phenomenon of overscreening is extremely sensitive to the anisotropy of the Kondo coupling and is only expected to be observed for isotropic coupling~\cite{Nozieres1995}. For an anisotropic case, the channel with the larger value of Kondo coupling would win over the other channel and form a singlet.

From Eqs.~\eqref{eq:RG_1} and~\eqref{eq:RG_2}, we can calculate the Kondo temperature for the two channels. We define $\alpha^j = \sqrt{({\tilde J}'_{z,j,0})^2/(J_{\perp,j,0})^2-1}$, where ${\tilde J}'_{z,j,0}, J_{\perp,j,0}$ are the bare values of the $\tilde{J}'_{z,j}, J_{\perp,j}$ couplings. With $\Lambda$ being the bandwidth of the original system, the Kondo temperature $T^j_{\rm K}$ for channel $j$ is given by 
\begin{equation} 
	T^j_{\rm K} = \Lambda \exp\Big( - 
	\frac{\sinh^{-1}( \alpha_j ) }{\alpha_j \nu_j J_{\perp,j,0} } \Big).
\end{equation}	
We note that $T^j_K$ is the characteristic energy scale of the Kondo effect pertaining to each of these channels. 
\section{Observable consequence}\label{ObsCon}
The Kondo effect gives rise to a negative correction, $\delta \mathcal{G}(\omega)$, to the conductance of the HLs, originating 
from spin-flip scattering processes mediated by the magnetic impurity. This correction vanishes for HLs as temperature $T$ $\to 0 $ and also in the DC limit when frequency $\omega \to 0$~\cite{Zhang2009,Furusaki2011}. 
However, the signature of Kondo scattering can still be captured by computing 
the correction at non-zero $T$ and $\omega$. In order to study the scaling of the 
correction to the conductance, one can compute $\delta \mathcal{G}(\omega)$ by incorporating a difference of chemical potential between the right and left movers, respectively. This is equivalent to computing $\delta \mathcal{G}(\omega) $ from the spin-flip current obtained by introducing an effective magnetic field which creates an energy difference between the spin up and the spin down components of the impurity~\cite{Furusaki2011,Eriksson2013}. We assume weak backscattering by the impurity such that $\delta \mathcal{G}(\omega) \ll e^2$ where $e$ is the electron charge. 

Following~\cite{Furusaki2011}, we attach $H_{\mathcal{V}}= -e\mathcal{V} \sigma^z $ to the Hamiltonian of Eq.~\eqref{Ham} and compute transport properties in response to the spin-flip current using $H_{\mathcal{V}}$. One can use the Kubo formula~\cite{bColeman,Furusaki2011,Eriksson2013} to compute this correction. The details of the calculation, including the general form of the correction, are given in App.~\ref{app-Cond}. One can write down the exact expressions for the conductance correction in the limit $J^2_{\perp,j} \ll \omega \ll T$. The temperature scaling of $\delta \mathcal{G}$ is given by
\begin{widetext}
	\begin{equation}
		\label{cond_LT} 
		\delta \mathcal{G} = \mathcal{L}_1 T^{2(\kappa^2_{11} + \kappa^2_{12})-2} + 
		\mathcal{L}_2 T^{2(\kappa^2_{21} + \kappa^2_{22})-2} + \mathcal{L}_3 T^{2(\kappa_{11}\kappa_{12} +
			\kappa_{22}\kappa_{21})-2},
	\end{equation}
	where
	\begin{align}
		\mathcal{L}_1 & = -  \frac{e^2}{4}
		\left( \frac{J_{\perp,1}  }{2 \pi \xi_1} \right)^2 
		\left( \frac{2\pi }{\Lambda} \right)^{2 \kappa^2_{11}}
		\left( \frac{2\pi }{\Lambda} \right)^{2 \kappa^2_{12}}
		\Big(\frac{1}{\pi }\Big)^2
		\frac{\pi \Gamma(\kappa^2_{11} + \kappa^2_{12})^2}{ \Gamma(2(\kappa^2_{11} + \kappa^2_{12}))}, 
		\\
		\mathcal{L}_2 & = - \frac{e^2}{4}
		\left( \frac{J_{\perp,2}  }{2 \pi \xi_2} \right)^2 
		\left( \frac{2\pi }{\Lambda} \right)^{2 \kappa^2_{21} } 
		\left( \frac{2\pi }{\Lambda} \right)^{2 \kappa^2_{22} }
		\Big(\frac{1}{\pi }\Big)^2
		\frac{\pi \Gamma(\kappa^2_{21} + \kappa^2_{22})^2}{ \Gamma(2(\kappa^2_{21} + \kappa^2_{22}))},  
		\\
		\mathcal{L}_3 & = - \frac{e^2}{2}
		\frac{J_{\perp,1} J_{\perp,2} }{(2 \pi)^2 \xi_1 \xi_2} 
		\left( \frac{2\pi }{\Lambda} \right)^{2(\kappa_{11}\kappa_{12})} 
		\left( \frac{2\pi }{\Lambda} \right)^{2\kappa_{22}\kappa_{21}} 
		\Big(\frac{1}{\pi }\Big)^2
		\frac{\pi \Gamma(\kappa_{11}\kappa_{12} +\kappa_{22}\kappa_{21})^2}{\Gamma(2(\kappa_{11}\kappa_{12} +\kappa_{22}\kappa_{21}))}.
	\end{align}	
\end{widetext}
The result takes a form similar to the known case of one HL coupled to a Kondo impurity~\cite{Zhang2009,Furusaki2011}. We notice that, at the exactly solvable point, the integral from which $\mathcal{L}_3$ is obtained, goes to zero, as discussed in App.~\ref{app-Cond} and one is left with $\mathcal{L}_1$ and $\mathcal{L}_2$. The assumed limit for the above expression is important, as we see that at high temperature (i.e. $T \gg \omega \gg J^2_{\perp,j} $) the Kondo scattering gives rise to a negative correction, implying deviation from standard scaling of the conductance in HLs in the absence of Kondo impurity. We also see that in this limit the correction is independent of $\omega$.

\section{Conclusion}\label{Sum}

In this paper, we have presented a general framework for studying 
two interacting helical liquids coupled to a Kondo impurity, 
including both intra-channel as well as inter-channel interactions.

We have derived the conditions under which an exact solution of the model can be obtained 
with the additional restriction of $\kappa_j=1/\sqrt{2}$, where $\kappa_j$ is defined in Eq.~\eqref{defkappa}. This solvable point has been calculated using an Emery-Kivelson  type of transformation. We have included inter-channel 
forward scattering processes yielding exact solutions to the problem examined 
beyond the scenarios captured by Refs.~\cite{Emery1992,Posske2013,Sengupta1994}. 
In Sec.~\ref{ExSol} we have shown how the exact solutions of the decoupled limit 
(obtained by switching off $g^{12}_{2,4}$) can be derived from our calculation. 
At the solvable point, we have calculated the spectral function. We have shown 
that the level width is the sum of contributions coming from each channel. 
The spectral function has experimental significance in many systems; for example, quantum dots show the Kondo effect where the hallmark of Kondo physics is the differential conductance which is proportional to the spectral function~\cite{Gossard2005,Lindelof2000,Delft2003,Glazman2001}.

We have studied the model away from the exactly solvable point, by mapping it to a pair of  
effectively decoupled HLs interacting with a single magnetic impurity, 
as derived in Eq.~\eqref{eq:H_2CK_ind}. By using a perturbative RG approach, 
we have shown that these two renormalized channels can separately flow to either the 
FM or the AFM fixed point. Here, the AFM fixed point indicates the formation of a bound state between the conduction electron and impurity spin, 
whereas the FM fixed point means the absence of the same, although a finite residual coupling 
can be present in the FM case. In Fig.~\ref{fig:RG} we show a schematic flow diagram 
pertaining to the RG equations derived in Sec.~\ref{weakcouplingRG}. The phase diagrams obtained from our RG analysis are shown in Fig.~\ref{fig:Phase}. As noted earlier, these phase diagrams represent the parameter space of the Hamiltonian where the effectively decoupled HLs reach different FPs in different segments of the diagram. In Sec.~\ref{weakcouplingRG}, we also have discussed the nature of the impurity screening at different fixed points. In the FF phase, there is no singlet formation due to the absence of screening. In the FA and AF phases, the impurity is screened and in the AA phase, the impurity is either screened or overscreened depending on anisotropy in Kondo couplings. In Sec.~\ref{ObsCon}, we have presented a study of the linear response of the system in the weak coupling limit. We have shown that the Kondo effect gives rise to a negative correction to the conductance of the HLs. The temperature scaling of this correction as a function of system parameters has been shown explicitly.

\section{Acknowledgement}
S.B. and A.K. acknowledges supports from the DST (Govt. of India) via sanction no. DST /PHY /2021083, the SERB (Govt. of India) via sanction no. ECR/2018/001443 and CRG/2020/001803, DAE (Govt. of India ) via sanction no. 58/20/15/2019-BRNS, as well as MHRD (Govt. of India) via sanction no. SPARC/2018-2019/P538/SL.

\appendix

\begin{widetext}

	\section{Unitary transformations} \label{app-DCh}
	Under the action of the unitary transformation 
	\begin{equation}
		\label{unitarytransfo}
		U = e^{i 2\sqrt{\pi}( \lambda_1 \tilde{\phi}_1(0) + \lambda_2 \tilde{\phi}_2(0) ) \sigma^z},
	\end{equation}
	the Luttinger liquid Hamiltonian  $H_{\rm LL}$ transforms as
	\begin{equation}
		H_{\rm LL} \rightarrow 
		\tilde{H}_{\rm LL}
		= U H_{\rm LL} U^{\dagger}  
		= \sum_{j=1,2} \left[ \frac{u_j}{2} \int dx \left[  \tilde{\Pi}_j^2 + ( \partial_x \tilde{\phi}_j)^2  \right]
		-  2\sqrt{\pi} \lambda_j  u_j \tilde{\Pi}_j(0)  \sigma^z \right],
	\end{equation}
	while the Kondo Hamiltonian $H_{\rm K}$ transforms as 
	\begin{equation}
		H_{\rm K} 
		\rightarrow 
		\tilde{H}_{\rm K} = U H_{\rm K} U^{\dagger} =\sum_{j=1,2}
		\left[
		-\frac{\tilde J_{z,j}} {\sqrt{\pi}} \tilde{\Pi}_j(0)  \sigma^{z} 
		+ \frac{J_{\perp,j} }{2 \pi \xi_j} 
		\left( e^{i 2 \sqrt{\pi} [(V^{j1}_{\phi}+ \lambda_1)\tilde{\phi}_1(0) 
			+ (V^{j2}_{\phi} +\lambda_2)\tilde{\phi}_2(0)]} \sigma^{+} 
		+ {\rm h.c.} 
		\right) \right].
	\end{equation}
	(We have omitted an unimportant constant.) We then arrive at a decoupled-channel  Hamiltonian with either of the two following choices:
	\begin{align}
		\lambda_j=-V_\phi^{jj} \quad  \text{or} \quad
		\lambda_j=-V_\phi^{\bar j j} .
	\end{align}
	(We use the notation $\bar j=2,1$ for $j=1,2)$. 
	We select the second option, and collecting $\tilde H_{\rm LL}$ and $\tilde H_{\rm K}$, we arrive at the Hamiltonian
	\begin{equation}
		\tilde{H} 
		= \sum_{j=1,2}\left[ 
		\frac{u_j}{2} \int dx \left[  \tilde{\Pi}_j^2 +  ( \partial_x \tilde{\phi}_j )^2  \right] 
		-\frac{ \tilde J'_{z,j} }{\sqrt{\pi}} 
		\tilde{\Pi}_j(0) \sigma^{z} 
		+ \frac{J_{\perp,j} }{2 \pi \xi_j} \left(
		e^{i 2 \sqrt{\pi} \kappa_j 
			\tilde{\phi}_j(0)} \sigma^{+} + {\rm h.c. }\right) \right],
	\end{equation}
	where we have defined 
	\begin{align}
		\label{newdef}
		\tilde  J'_{z,j} = \tilde J_{z,j} 
		-  2\pi u_j  V^{\bar jj}_{\phi} , 
		\quad \kappa_j &= V^{jj}_{\phi}-V^{\bar jj}_{\phi}.
	\end{align}
	We use this Hamiltonian for the calculation of the RG flow of the Kondo couplings in Sec.~\ref{weakcouplingRG} and 
	App.~\ref{app-RG}. 
	
	Alternatively, in the unitary transformation~\eqref{unitarytransfo} 
	we can set  $\lambda_j=-\frac{\tilde J_{z,j}}{2\pi u_j}$ and we obtain
	\begin{align}
		\tilde{H} 
		& = \sum_{j=1,2}\left[ 
		\frac{u_j}{2} \int dx \left[  \tilde{\Pi}_j^2 +  ( \partial_x \tilde{\phi}_j )^2  \right] 
		+ \frac{J_{\perp,j} }{2 \pi \xi_j} \left(
		e^{i 2 \sqrt{\pi} \sum_k \kappa_{jk}\tilde{\phi}_k(0)} \sigma^{+} + {\rm h.c. }\right) \right],
	\end{align}	
	where we have defined 
	\begin{equation}
		\kappa_{jk}=V_\phi^{jk} - 
		\frac{\tilde J_{z,k}}{2\pi u_k}.
	\end{equation}
	We employ this form of the Hamiltonian in the discussion of the solvable point in Sec.~\ref{ExSol} and for the perturbative calculation  of the correction to the conductance in Sec.~\ref{ObsCon}. 
	\section{Interaction Hamiltonian} \label{app-BSDim}
	In this appendix we establish the conditions under which it is justified to retain only  the interaction terms included in Eq.~\eqref{Hint}. Following \cite{Nagaosa2009}, the interaction Hamiltonian for a two-channel HL in general 
	comprises terms that lead to nonlinearities in the bosonized theory. These terms include the Umklapp scattering processes
	\begin{align}
		&\Psi^{\dagger}_{j \uparrow}(x) \Psi^{\dagger}_{j \uparrow}(x+a) 
		\Psi_{j \downarrow}(x+a) \Psi_{j \downarrow}(x) e^{-i4k_Fx}
		+{\rm h.c.}, \quad j=1,2.
	\end{align}
	We omit these processes on account of the fact that we consider 
	the generic incommensurate situation, i.e., $4k_F$ different from a reciprocal lattice vector.
	
	Next, let us consider 
	Kondo scatterings between different channels. They are described by the following operators:
	\begin{align}
		\Psi^{\dagger}_{1,\uparrow} \Psi_{2,\uparrow} - \Psi^{\dagger}_{1,\downarrow} \Psi_{2,\downarrow} 
		&\sim 
		e^{i\sqrt{\pi}( (V^{11}_{\theta}- V^{21}_{\theta}) \tilde{\theta}_1 + (V^{12}_{\theta} - V^{22}_{\theta} )\tilde{\theta}_2 )} \sin(\sqrt{\pi}( (V^{11}_{\phi}- V^{21}_{\phi}) \tilde{\phi}_1 + (V^{12}_{\phi}  - V^{22}_{\phi}) \tilde{\phi}_2 )), \\
		\Psi^{\dagger}_{1,\uparrow}  \Psi_{2,\downarrow} 
		& \sim 
		e^{i\sqrt{\pi}( (V^{11}_{\theta}- V^{21}_{\theta}) \tilde{\theta}_1 + 
			(V^{12}_{\theta} - V^{22}_{\theta} )\tilde{\theta}_2 )} 
		e^{-i\sqrt{\pi}( (V^{11}_{\phi}+ V^{21}_{\phi}) \tilde{\phi}_1 + 
			(V^{12}_{\phi} + V^{22}_{\phi} )\tilde{\phi}_2 )} .
	\end{align}
	Calculating their scaling dimensions, 
	we find that these terms are irrelevant (and can thus be omitted) if the following two conditions hold:
	\begin{equation}
		\frac{1}{4} \sum_{j=1,2} \left(V_\theta^{1j}-V_\theta^{2j} \right)^2 
		+ \frac{1}{4} \sum_{j=1,2} \left(V_\phi^{1j}
		\mp V_\phi^{2j}\right)^2 > 1.
	\end{equation}
	\section{Two-channel resonant-level model} \label{app-2RLM}
	In this section, we briefly discuss the solution of Eq.~\eqref{HT}. We Fourier transform the Hamiltonian in this equation to cast it into a resonant-level model consisting of two non-interacting channels coupled to a discrete level modeled by $d$ operators. The Hamiltonian becomes 
	\begin{equation}
		\label{appHrm}
		H_{\rm T}
		= \sum_{k, j} \epsilon_{k,j} c^{\dagger}_{k,j} c_{k,j} + \epsilon_d d^{\dagger} d 
		+ \sum_{k,J} t_{j}\Big[  c^{\dagger}_{k,j} d + c_{k,j} d^{\dagger} \Big].
	\end{equation}
	Here, $t_j = \frac{J_{\perp,j}  }{\sqrt{4 \pi^2 \xi_j}}$ and $\epsilon_{k,j} =  u_{j} k $. We look for new fermionic operators $f^{\dagger}_{n} = 
	\sum_{k,j} M^{j}_{n,k} c^{\dagger}_{k,j} + L_n d^{\dagger}$ such that $H_{\rm T}=\sum_n E_n f_n^{\dagger}f_n + {\rm const}$. We then have $[ H_{\rm T},f^{\dagger}_n ]= E_nf_n^{\dagger}$, and from Eq.~\eqref{appHrm}
	\begin{equation}
		\begin{aligned}
			\Big[ H_{\rm T} , f^{\dagger}_n \Big]
			=   \sum_{k,j} [M^{j}_{n,k} \epsilon_{k,j} c^{\dagger}_{k,j} +  t_{j}M^{j}_{n,k}  d^{\dagger} + t_{j} L_{n} c_{k,j}^{\dagger} ]  + \epsilon_d L_n  d^{\dagger} .
		\end{aligned}	
	\end{equation}
	We have used $[ab,c] = a\{b,c\} - \{a,c\}b$.  Hence
	\begin{equation}
		\begin{aligned}
			E_{n} f_n^{\dagger} 
			& = \sum_{k,j} [M^{j}_{n,k} \epsilon_{k,j} c^{\dagger}_{k,j} +  t_{j}M^{j}_{n,k}  d^{\dagger} +  t_{j} L_{n} c_{k,j}^{\dagger} ]  + \epsilon_d L_n  d^{\dagger}  ,\\
			E_{n} \Big[ \sum_{k,j}  M^{j}_{n,k} c^{\dagger}_{k,j} + L_n d^{\dagger} \Big] & = \sum_{k,j} [ M^{j}_{n,k} \epsilon_{k,j} c^{\dagger}_{k,j} +    t_{j} L_{n} c_{k,j}^{\dagger} ]  + [\sum_{k,j}  t_{j}  M^{j}_{n,k}  d^{\dagger} +\epsilon_d L_n  d^{\dagger}] .
		\end{aligned}	
	\end{equation}
	If we introduce a resonant level coupled to a two independent channel set-up, then
	\begin{equation}
		\begin{aligned}
			& E_{n} M^{ j}_{n,k}  =  M^{ j}_{n,k} \epsilon_{k, j} +  t_{ j} L_{n}  ,\\
			& E_{n} L_{n} = \sum_{ j}  t_{ j} \sum_k M^{j}_{n,k}  + \epsilon_d L_{n} .
		\end{aligned}	
	\end{equation}
	One would get
	\begin{equation}
		\begin{aligned}
			E_{n}  & = \sum_{j} t^2_{ j} \sum_k \frac{1}{E_{n}-\epsilon_{k,j}} + \epsilon_d .
		\end{aligned}	
	\end{equation}
	The above equation can be graphically solved for finite system \cite{bColeman}. We use the following relation \cite{bColeman} to evaluate the sum over $k$
	\begin{equation}
		\begin{aligned}
			\sum_{n= -\infty}^{\infty} \frac{1}{E_n - \pi n } = \cot(E_n) .
		\end{aligned}	
	\end{equation}
	For $\epsilon_d=0$ 
	\begin{equation}
		\begin{aligned}
			E_{n}  
			& =
			\sum_{j} \frac{\pi t^2_{j}}{u_j} \cot(\frac{E_{n} \pi}{u_j}) .
		\end{aligned}	
	\end{equation}
	One can solve the above equation numerically to obtain the energy $E_n$. We can further derive the spectral function from the impurity Green's function. We note that in path integral formalism
	\begin{equation}
		\begin{aligned}
			Z &= \int \mathcal{D}[d,c_j] e^{-S} = \int \mathcal{D}[d] e^{-S_d} \int \mathcal{D}[c] e^{-S_{c}} ,\\
			S_d &= \int_{0}^{\beta} d \tau \Big[ \bar{d} (\partial_\tau + \epsilon_d) d \Big]  ,\\
			S_c &= \sum_j \int_{0}^{\beta} d \tau \Big[ \sum_k \bar{c}_{k,j} (\partial_\tau + \epsilon_{k,j} ) c_{k,j} +  \sum_{k} ( t_{j} \bar{c}_{k,j} d + c_{k,j} \bar{d}  )\Big] .\\
		\end{aligned}	
	\end{equation}  
	From the above expressions we can write
	\begin{equation}
		\begin{aligned}
			Z = \int \mathcal{D} d e^{-\int_{0}^{\beta} d \tau \Big[ \bar{d} (\partial_\tau + \epsilon_d) d \Big]} &\int \mathcal{D}[c_1] e^{-\int_{0}^{\beta} d \tau \Big[ \sum_k \bar{c}_{k,1} (\partial_\tau + \epsilon_{k,1} ) c_{k,1} +  \sum_{k} ( t_{\alpha} \bar{c}_{k,1} d + c_{k,1} \bar{d}  )\Big]} \\
			& \int \mathcal{D}[c_2] e^{-\int_{0}^{\beta} d \tau \Big[ \sum_k \bar{c}_{k,2} (\partial_\tau + \epsilon_{k,2} ) c_{k,2} +  \sum_{k} ( t_{\alpha} \bar{c}_{k,2} d + c_{k,2} \bar{d}  )\Big]}  .
		\end{aligned}	
	\end{equation}  
	One can integrate out $c_{k,j}$'s to obtain
	\begin{equation}
		\begin{aligned}
			Z \sim
			\int \mathcal{D} d \exp\Big[-\int_{0}^{\beta} d \tau \bar{d} (\partial_\tau + \epsilon_d - \sum_j \frac{t^2_j}{\partial_\tau + \epsilon_{k,j}}) d \Big ].
		\end{aligned}	
	\end{equation}  
	Here we are showing the relevant term which depends on $d$ operators. We next perform Fourier transformation $d(\tau) = \beta^{-1/2} \sum_n d_n e^{-i \omega_n \tau}$, where $\omega_n$ is $n^{th}$ Matsubara frequency and $\beta$ is inverse temperature. This enables us to replace $\partial_\tau$ by $- i \omega_n$. 
	\begin{equation}
		\begin{aligned}
			Z \sim
			\int \mathcal{D} d \exp\Big[- \sum_{i \omega_n} \bar{d_n} (- i \omega_n + \epsilon_d - \sum_j \frac{t^2_j}{-i \omega_n + \epsilon_{k,j}}) d_n \Big ].
		\end{aligned}	
	\end{equation} 
	Hence the impurity green's function can be written as
	\begin{equation}
		\begin{aligned}
			G_d(i\omega_n) = \frac{1}{i \omega_n -\epsilon_d + i\Gamma sgn (\omega_n)}.
		\end{aligned}	
	\end{equation} 
	and the spectral function, defined as $-\frac1\pi $Im($G_d$), becomes a Lorentzian with width
	\begin{equation}
		\begin{aligned}
			\Gamma = \sum_j \Gamma_j = \sum_j \pi \frac{t^2_j}{u_j} .
		\end{aligned}	
	\end{equation}
	\section{RG analysis} \label{app-RG}
	For completeness, in this appendix we provide the derivation of the flow equations for the Kondo couplings in the model \eqref{eq:H_2CK_ind} using the perturbative RG approach. (See, e.g., \cite{maciejko2012}). The (euclidean) action for the Kondo problem corresponding to the Hamiltonian~\eqref{eq:H_2CK_ind} is $S=S_0+S_{\rm K}$, where
	\begin{align}
		S_0 & = \sum_{j=1,2} \int \frac{d\omega}{2\pi} |\omega|
		|\varphi_j(\omega) |^2,\\
		S_{\rm K}
		& = \sum_{j=1,2} \int d\tau \left[ \frac{-i\tilde J'_{z,j}}{\sqrt{\pi} u_j}\partial_\tau  \varphi_j \, \sigma^{z} +  \frac{J_{\perp,j} }{2 \pi \xi_j}	\left( e^{i 2 \sqrt{\pi} \kappa_j  \varphi_j} \sigma^{+} + 
		e^{-i 2 \sqrt{\pi} \kappa_j \varphi_j} \sigma^{-} \right) \right].
	\end{align}	
	Here we use the notation $\varphi_j(\tau) \equiv \tilde \phi_j(0,\tau)$, and $\tau$ denotes imaginary time. Following \cite{maciejko2012} $S_0$ is obtained by integrating out $\varphi_j(x \neq 0,\tau)$. The action $S$ contains a large frequency cutoff $\Lambda$, i.e., $|\omega| < \Lambda$ in all frequency integrations, where we identify
	$\Lambda=\frac{u_j}{\xi_j}$.
	
	The RG approach proceeds as follows \cite{maciejko2012,bAltland}. We introduce a rescaled cutoff $\Lambda'= \Lambda/b$, where $b=e^l>1$ is a scaling factor, with $\ell\ll 1$. We separate the field into slow and fast components, $\varphi_j = \varphi_j^< + \varphi_j^>$, where the first contains only frequency components smaller than $\Lambda'$, and the latter contains frequency components between $\Lambda'$ and $\Lambda$. We perform the same separation on $\vec{\sigma}$ as we do for the $\varphi_j$ fields. We are using time-ordered bosonic correlation, hence for consistency of the calculation one has to use the time-ordered product of impurity spin as well. This is given by $\mathcal{T} [\sigma^{\pm}_<(\tau) \sigma^{\mp}_<(\tau')] = \frac12 + \sigma^z_{<} \text{sgn}\,(\tau-\tau')$ \cite{maciejko2012}. We then integrate over the fast component and obtain an effective action for the slow component, which has the same form as the original action, but with renormalized coefficients, from which we can read the RG equations. 
	
	After integrating out the fast modes, the effective action for the slow modes 
	up to second order in the Kondo couplings takes the following expression:
	\begin{align}
		S_{\rm eff} [\varphi^<]
		& =
		S_0[\varphi^< ] +
		\langle S_{\rm K}[\varphi] \rangle_{>} - \frac12 
		\left( \langle S^2_{\rm K}[\varphi] \rangle_{>} - \langle S_{\rm K}[\varphi] \rangle^{2}_{>} \right), 
	\end{align}	
	where $\langle \dots \rangle_>$ denotes the integration over the fast modes, i.e., using the action $S_0[\varphi^>]$. Let us begin with the calculation of the first-order correction. Here, we have used $ \langle \partial_\tau \varphi_j \rangle_{>} =  \partial_\tau \varphi_j^{<}$ 
	and $\langle (\varphi_j^{>} )^2 \rangle_{>} = \frac{1}{2\pi} \log b$. We now need to restore the cutoff to its original value $\Lambda$, which can be accomplished by rescaling the time $\tau \rightarrow b\tau$ and redefining the field $\varphi_j^<(\tau) \rightarrow \varphi_j^<(b\tau) \equiv\bar \varphi(\tau)$. Then we get 
	\begin{align}
		\langle S_{\rm K} [\varphi]\rangle_{>}
		=  \sum_j \int d \tau
		\left[ \frac{-i \tilde J'_{z,j}}{\sqrt{\pi} u_j} 
		\partial_\tau \bar \varphi_j \, \sigma^{z} +
		\frac{J_{\perp,j} b^{1-\kappa_j^2}}{2 \pi \xi_j } 
		\left( 
		e^{i 2 \sqrt{\pi} \kappa_j \bar \varphi_j } \sigma^+ + 
		e^{-i 2 \sqrt{\pi} \kappa_j \bar \varphi_j} \sigma^-
		\right) \right].
	\end{align}	
	This results implies that $\tilde{J}'_{z,j}$ is not renormalized at first order, while for $J_{\perp,j}$ we find $$ \frac{dJ_{\perp,j}}{d\ell} = (1-\kappa_j^2) J_{\perp,j}. $$
	At the second order we find 
	\begin{align}
		\langle S^2_{\rm K}[\varphi] \rangle_{>} 
		&=  \sum_{j_1,j_2}\int d\tau\,d\tau'  
		\Big\langle
		\left[
		\frac{-i \tilde J'_{z,j_1} }{\sqrt{\pi}u_{j_1}} \partial_\tau \varphi_{j_1} (\tau) \sigma^{z} +
		\frac{J_{\perp,j_1} }{2 \pi \xi_{j_1}} 
		\left( 
		e^{i 2 \sqrt{\pi} \kappa_{j_1} \varphi_{j_1}(\tau)} \sigma^{+} + 
		e^{-i 2 \sqrt{\pi} \kappa_{j_1} \varphi_{j_1}(\tau)} \sigma^{-} 
		\right)  \right] \times \nonumber  \\
		&\times \left[
		\frac{-i \tilde J'_{z,j_2} }{\sqrt{\pi}u_{j_2}} \partial_{\tau'}\varphi_{j_1}(\tau') \sigma^{z} + 
		\frac{J_{\perp,j_2}  }{2 \pi \xi_{j_2}} 
		\left( 
		e^{i 2 \sqrt{\pi} \kappa_{j_2} \varphi_{j_2} (\tau')} \sigma^{+} + 
		e^{-i 2 \sqrt{\pi} \kappa_{j_2} \varphi_{j_2} (\tau')} \sigma^{-} 
		\right)  \right] \Big\rangle_> .
	\end{align}	
	We calculate only the quantity of interest:
	\begin{align} \Big\langle
		e^{i 2 s \sqrt{\pi} \kappa_j \varphi_j (\tau)} \partial_{\tau'} \varphi_j^{>}(\tau') \Big\rangle_{>} 
		&=
		-\frac{2 s \sqrt{\pi}}{i b^{\kappa_j}} e^{i 2 s \sqrt{\pi} \kappa_j \varphi_j^{<}(\tau)} \kappa_j  \partial_{\tau'} \langle \varphi_j^{>}(\tau) \varphi_j^{>}(\tau') \rangle_{>} .
	\end{align}	
	Next, we make the change of variables $t = \tau - \tau'$ and $T=\frac{\tau+\tau'}{2}$, and 
	Taylor-expand the fields around $t=0$, 
	so that they are only functions of the center of mass coordinate $T$. The $t$-integral is performed using a cut-off $1/\Lambda$.
	Collecting together the dominant terms, we obtain
	\begin{align}
		\langle S_{\rm K} \rangle_{>}
		& =   \int dT \left[ \frac{-i \tilde{J}'_{z,j} }{\sqrt{\pi}u_j} 
		\partial_{T} \bar{\varphi_j}(T) \sigma^{z} +
		\frac{J_{\perp,_j}  }{2 \pi \xi} 
		\left( 1 + (\kappa_j^2-1)\frac{\delta \Lambda}{\Lambda}\right) 
		\left( 
		e^{i 2 \sqrt{\pi}\kappa_j \bar{\varphi_j}(T)} \sigma^{+} + 
		e^{-i 2 \sqrt{\pi}\kappa_j \bar{\varphi_j}(T)} \sigma^{-} 
		\right) \right]  ,\nonumber  \\
		-\frac12 ( \langle S^2_{\rm K} \rangle_{>} - \langle S_{\rm K} \rangle^{2}_{>} )
		& =
		\sum_{j=1}^{2} \kappa_j \int dT  
		\frac{J_{\perp,j} \tilde{J}'_{z,_j} }{2 \pi^2 \xi_j u_j}   \log b
		\Bigg[~ e^{i 2 \sqrt{\pi} \kappa_j \varphi^{<}_j (T) }  \sigma^{+}
		+e^{-i 2 \sqrt{\pi} \kappa_j \varphi^{<}_j (T) }  \sigma^{-} \Bigg]  \nonumber  \\ 
		& -  \sum_{j=1}^{2}  \frac{(J_{\perp,j})^2 }{(2 \pi)^2 \xi^2_j}  4 \sqrt{\pi}   \kappa^3_j i\sigma^z \int dT  \partial_T\varphi_j^{<}(T) \frac{1}{\Lambda^2}  dl .
	\end{align}	
	We then arrive at the effective action with rescaled couplings
	\begin{align}
		S_{\rm eff}[\bar \varphi] = S_0[\bar \varphi] +
		\int dT 
		\left[ \frac{-i\tilde J'_{z,j}(l) }{\sqrt{\pi}u_j} 
		\partial_T \bar{\varphi}_j(T) \sigma^{z} +
		\frac{J_{\perp,j}(l)}{2 \pi \xi_j} 
		\left( 
		e^{i 2 \sqrt{\pi}\kappa_j \bar{\varphi}_j(T)}
		\sigma^{+} + 
		e^{-i 2 \sqrt{\pi}\kappa_j \bar{\varphi}_j(T)} \sigma^{-} 
		\right) \right].
	\end{align}	
	where the couplings satisfy the following RG equations:
	\begin{align}
		\frac{d \tilde J'_{z,j}}{dl} 
		&=  
		\frac{  \kappa^3_j }{ \pi u_j} (J_{\perp,j})^2,\label{RGJz}\\
		\frac{d J_{\perp,j}}{dl}
		&=
		(1-\kappa^2_j)J_{\perp,j}
		+
		\frac{\kappa_j}{ \pi u_j } \tilde J'_{z,j} J_{\perp,j} . \label{RGJperp}
	\end{align}	
	In terms of the unshifted couplings $\tilde J_{z,j}$, we find
	\begin{align}
		\frac{d \tilde J_{z,j}}{dl} 
		&=  
		\frac{  \kappa^3_j }{ \pi u_j} (J_{\perp,j})^2,\\
		\frac{d J_{\perp,j}}{dl}
		&=
		(1-\kappa^2_j - 2 \kappa_j V_\phi^{\bar j j} ) J_{\perp,j}
		+
		\frac{\kappa_j}{ \pi u_j } \tilde J_{z,j} J_{\perp,j} ,\nonumber \\
		& = (1-(V_\phi^{jj})^2 - (V_\phi^{\bar j j})^2 ) J_{\perp,j}
		+
		\frac{\kappa_j}{ \pi u_j } \tilde J_{z,j} J_{\perp,j}.
	\end{align}	
	The first term in the last line can be easily understood: it is the scaling dimension of the $J_\perp$ operator in Eq.~\eqref{HKcoupled}, after the diagonalization of the bulk Hamiltonian, but before any unitary transformation. In the limit of decoupled channels, 
	$$
	V_\phi^{ij} = \sqrt{K_j} \delta_{ij}, \quad V_\theta^{ij} =
	\delta_{ij}/ \sqrt{K_j},\quad \tilde J'_{z,j} = J_{z,j}/\sqrt{K_j},\quad \kappa_j =\sqrt{K_j},
	$$
	and we find
	\begin{align}
		\frac{d J_{z,j}}{dl} 
		&=  
		\frac{  K^2_j }{ \pi u_j} (J_{\perp,j})^2,\\
		\frac{d J_{\perp,j}}{dl}
		&=
		(1-K_j)J_{\perp,j}
		+
		\frac{1}{ \pi u_j } J_{z,j} J_{\perp,j} .
	\end{align}	
	The equations differ from those in \cite{Zhang2009}, which read
	\begin{align}
		\frac{d J_{z,j}}{dl} 
		&=  
		\frac{  1}{ \pi u_j} (J_{\perp,j})^2,\\
		\frac{d J_{\perp,j}}{dl}
		&=
		(1-K_j)J_{\perp,j}
		+
		\frac{1}{ \pi u_j } J_{z,j} J_{\perp,j} .
	\end{align}
	This can be understood as being due to the fact that the equations in \cite{Zhang2009} are only meant  to be valid for weak e-e interactions, i.e., $K_j\approx 1$ (see comment in \cite{maciejko2012}).
	
	We notice that the equation for $J_{\perp,j}$ can be easily obtained by using the Hamiltonian~\eqref{Ham}. Indeed, in this case we just need the scaling dimension of the $J_\perp$ operator:
	\begin{equation}\label{RGscalingJperp}
		\frac{dJ_{\perp,j}}{d\ell} = ( 1-\sum_k \kappa^2_{jk})J_{\perp,j}.
	\end{equation}
	We observe that 
	\begin{align}
		\sum_k \kappa^2_{jk} &= \sum_k \left( V_\phi^{jk} - \frac{\tilde J_{z,k}}{2\pi u_k} \right)^2=
		\sum_k \left( V_\phi^{jk} - \frac{\tilde J'_{z,k}}{2\pi u_k} -V^{\bar k k}_\phi \right)^2 ,\\
		&\approx
		\left( V_\phi^{jj}  -V^{\bar j j}_\phi \right)^2 - \left( V_\phi^{jj}  -V^{\bar jj}_\phi \right)^2 
		\frac{\tilde J'_{z,j}}{\pi u_j} = \kappa_j^2 - \frac{\kappa_j }{\pi u_j} \tilde J'_{z,j}.
	\end{align}
	In the last line we used the definition of $\tilde J'_{z,j}$ in Eq.~\eqref{defkappa} and we omitted terms of order $\tilde J'^2_z$. Inserting this expression in Eq.~\eqref{RGscalingJperp} we recover Eq.~\eqref{RGJperp}.
	\section{Conductance for weak coupling}  
	\label{app-Cond}
	
	In this section we give some details of the calculation of the correction to the conductance 
	at finite temperature and frequency. 
	We start with the Hamiltonian $\tilde H$ in Eq.~\eqref{Ham}, 
	which we rewrite here for convenience:
	\begin{align}
		\tilde{H} 
		& = \sum_{j=1,2} \left[ 
		\frac{u_j}{2}\int dx 
		\left[ \tilde{\Pi}_j^2 + ( \partial_x \tilde \phi_j )^2 \right] 
		+ \frac{J_{\perp,j} }{2 \pi \xi_j} \left( e^{i 2 \sqrt{\pi} 
			\sum_k \kappa_{jk} \tilde \phi_k(0)} \sigma^{+} + {\rm h.c.} \right) \right].
	\end{align}	
	The spin flip current is given by $\delta I = -e \partial_t \sigma^z$ \cite{Furusaki2011}. 
	In our case this expression turns out to be
	\begin{equation}
		\label{Spinflipcurrent}
		\delta I  = ie\sum_{j=1,2}
		\frac{J_{\perp,j}  }{2 \pi \xi_j} \left[ e^{i 2 \sqrt{\pi} \sum_k \kappa_{jk} \tilde \phi_k }  \sigma^+
		- e^{-i 2 \sqrt{\pi} \sum_k \kappa_{jk} \tilde \phi_k  } \sigma^{-} \right]  .
	\end{equation}
	The correction to the conductance $\delta \mathcal{G}(\omega)$ can be computed 
	using the Kubo formula \cite{bColeman,Eriksson2013}, which amounts to calculating the current-current correlator from Eq.~\eqref{Spinflipcurrent} \cite{Furusaki2011,Eriksson2013}. For these calculations one needs to use the finite temperature bosonic correlators defined as \cite{bGiamarchi,bFradkin}
	\begin{align}
		\langle \mathcal{T} \left[ \tilde \phi_j (\tau) - \tilde \phi_j(0) \right]^2 \rangle 
		& =\frac{1}{2 \pi} \log 
		\left[ \left(  \frac{\beta u_j}{\pi \xi_j} \right)^2
		\sin^2 \left( \frac{\pi}{\beta} \tau \right)   ~\right],
	\end{align}	
	where $\tau$ is imaginary time, $\beta = 1/T$, and $\xi_j = u_j/\Lambda$. 
	We then obtain $\delta \mathcal{G}(\omega) = I_1 + I_2 + I_3$, where
	\begin{equation}
		\begin{aligned}\label{cond}
			I_1 & = - 2 e^2
			\left( \frac{J_{\perp,1}  }{2 \pi \xi_1} \right)^2 
			\left( \frac{\pi \xi_1}{\beta u_1} \right)^{2 \kappa^2_{11}}
			\left( \frac{\pi \xi_2}{\beta u_2} \right)^{2 \kappa^2_{12}}
			\sin \left( \pi( \kappa^2_{11} + \kappa^2_{12}) \right) 
			\int_{0}^{\infty} dt  
			\frac{(e^{i \omega t}-1)/i\omega}{	| \sinh  (\frac{\pi}{\beta} t )  |^{  2(\kappa^2_{11} + \kappa^2_{12}) } }, 
			\\
			I_2 & = -2 e^2
			\left( \frac{J_{\perp,2}  }{2 \pi \xi_2} \right)^2 
			\left( \frac{\pi \xi_1}{\beta u_1} \right)^{2 \kappa^2_{21} } 
			\left( \frac{\pi \xi_2}{\beta u_2} \right)^{2 \kappa^2_{22} }
			\sin\left( \pi(\kappa^2_{21} + \kappa^2_{22} ) \right) 
			\int_{0}^{\infty} dt  
			\frac{(e^{i \omega t}-1)/i\omega}{| \sinh \left(\frac{\pi}{\beta} t\right)|^{2(\kappa^2_{21} + \kappa^2_{22}) }},  
			\\
			I_3 & = -4 e^2
			\frac{J_{\perp,1} J_{\perp,2} }{(2 \pi)^2 \xi_1 \xi_2} 
			\left( \frac{\pi \xi_1}{\beta u_1} \right)^{2(\kappa_{11}\kappa_{12})} 
			\left( \frac{\pi \xi_2}{\beta u_2} \right)^{2\kappa_{22}\kappa_{21}} 
			\sin\left(\pi( \kappa_{11}\kappa_{12} + \kappa_{22}\kappa_{21} ) \right) 
			\int_{0}^{\infty} dt  
			\frac{(e^{i \omega t}-1)/i\omega}{| \sinh \left( \frac{\pi}{\beta} t \right)  |^{2(\kappa_{11}\kappa_{12} +\kappa_{22}\kappa_{21}) } }  .
		\end{aligned}	
	\end{equation}
	
	In the limit $J^2_{\perp,j} \ll \omega \ll T$ we can simplify the above expressions as follows:
	\begin{equation}
		\begin{aligned}\label{CondI3}
			I_1 & = -  \frac{e^2}{4}
			\left( \frac{J_{\perp,1}  }{2 \pi \xi_1} \right)^2 
			\left( \frac{2\pi T}{\Lambda} \right)^{2 \kappa^2_{11}}
			\left( \frac{2\pi T}{\Lambda} \right)^{2 \kappa^2_{12}}
			\Big(\frac{1}{\pi T}\Big)^2
			\frac{\pi \Gamma(\kappa^2_{11} + \kappa^2_{12})^2}{ \Gamma(2(\kappa^2_{11} + \kappa^2_{12}))}, 
			\\
			I_2 & = - \frac{e^2}{4}
			\left( \frac{J_{\perp,2}  }{2 \pi \xi_2} \right)^2 
			\left( \frac{2\pi T}{\Lambda} \right)^{2 \kappa^2_{21} } 
			\left( \frac{2\pi T}{\Lambda} \right)^{2 \kappa^2_{22} }
			\Big(\frac{1}{\pi T}\Big)^2
			\frac{\pi \Gamma(\kappa^2_{21} + \kappa^2_{22})^2}{ \Gamma(2(\kappa^2_{21} + \kappa^2_{22}))},  
			\\
			I_3 & = - \frac{e^2}{2}
			\frac{J_{\perp,1} J_{\perp,2} }{(2 \pi)^2 \xi_1 \xi_2} 
			\left( \frac{2\pi T}{\Lambda} \right)^{2(\kappa_{11}\kappa_{12})} 
			\left( \frac{2\pi T}{\Lambda} \right)^{2\kappa_{22}\kappa_{21}} 
			\Big(\frac{1}{\pi T}\Big)^2
			\frac{\pi \Gamma(\kappa_{11}\kappa_{12} +\kappa_{22}\kappa_{21})^2}{\Gamma(2(\kappa_{11}\kappa_{12} +\kappa_{22}\kappa_{21}))}.
		\end{aligned}	
	\end{equation}
	The reduction of each of the integrals in Eq.~\eqref{cond} to the three corresponding expressions in Eq.~\eqref{CondI3} also depends on the fact that the terms $\kappa^2_{11} + \kappa^2_{12},\kappa^2_{21} + \kappa^2_{22} , \kappa_{11}\kappa_{12} + \kappa_{22}\kappa_{21} $ are non-zero. One can observe from Eq.~\eqref{cond} that these three terms correspond to $I_1, I_2$ and $I_3$ respectively. If any of these three quantities becomes zero, the corresponding integral goes to zero. We extend this argument to see from Eq.~\eqref{cond} that $I_3$ vanishes in the exactly solvable limit.
	This shows that in this limit 
	the correction takes the form of a sum of two contributions coming from two effectively independent channels, 
	as in the case of the level width of the spectral function in Sec.~\ref{ExSol}.
\end{widetext}

\end{document}